\documentclass[epj]{svjour}
\usepackage{graphics}
\newcommand{\beq}{\begin{equation}}
\newcommand{\eeq}{\end{equation}}
\begin{document}
\title{Wave-Packet Analysis of Single-Slit Ghost Diffraction}
\author{Tabish Qureshi\inst{1} \and Sheeba Shafaq\inst{2}}

\institute{Centre for Theoretical Physics, Jamia Millia Islamia, New Delhi,
\email{tabish@ctp-jamia.res.in}
\and 
Department of Physics, Jamia Millia Islamia, New Delhi,
\email{shafaqsheeba1@gmail.com}}

\abstract{
We show that single-slit two-photon ghost diffraction can be explained
very simply by using a wave-packet evolution of a generalised EPR state. 
Diffraction of a wave travelling in the x-direction can be described in
terms of the spreading in time of the transverse (z-direction) wave-packet,
within the Fresnel approximation. The slit is assumed to truncate the
transverse part of the wavefunction of the photon to within the width of
the slit. The analysis reproduces all features of
the two-photon single-slit ghost diffraction.  }


\maketitle

%\keywords{Entanglement; Nonlocality; Ghost interference.}

\section{Introduction}

The issue of quantum nonlocality which results from entangled states,
has been a subject of debate since the time it was brought to prominence
by the seminal paper of Einstein, Podolsky and Rosen (EPR)\cite{epr}.
Quantum systems which are said to be entangled, show certain correlation
in their measurement results even though they may be far separated in
space,\cite{entanglement}. This is a feature of quantum mechanics which
many find discomforting.

A dramatic demonstration of quantum 
nonlocality is in the so-called ghost interference experiment by
Strekalov et.al.\cite{ghostexpt}. We describe below, the single-slit
ghost diffraction experiment.
\begin{figure}[pb]
\centerline{\resizebox{13.0cm}{!}{\includegraphics{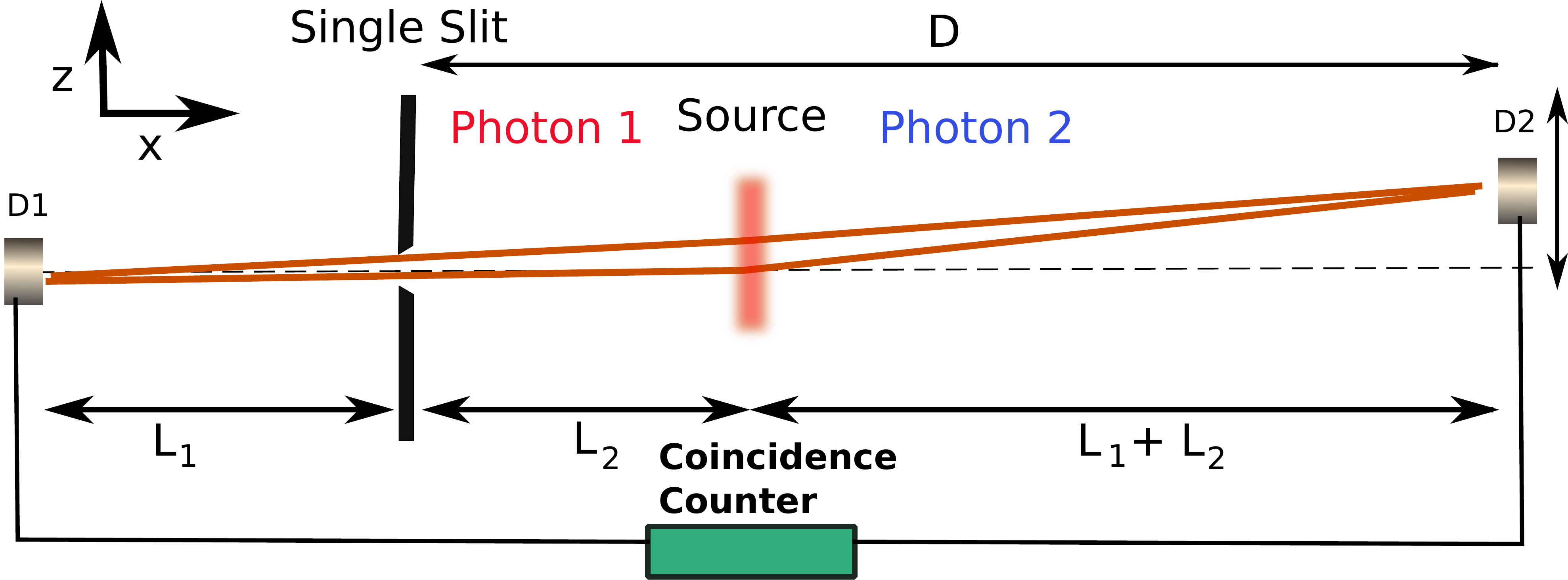}}}
\vspace*{8pt}
\caption{Schematic diagram of the single-slit ghost interference experiment.
Entangled photons 1 and 2 emerge from the SPDC source S and travel in
different directions along x-axis.}
\label{ghostexpt}
\end{figure}
An entangled pair of photons emerges from a spontaneous parametric down-conversion
(SPDC) source S. Photons, which we shall refer to as 1 and 2, are separated by
a beam-splitter and travel in different directions,
A single-slit is kept in the path of photon 1, and
there is a movable detector D1 behind it (see Fig. \ref{ghostexpt}).
Normally one would expect a single-slit diffraction pattern. No first order 
diffraction is observed for photon 1.  This result is unexpected because
one would expect photons from a laser, passing through a single-slit, to
undergo diffraction.
The second, and the more surprising result of the
experiment is that when photon 2 is detected by detector D2,
{\em in coincidence} with a {\em fixed} detector D1 detecting photon 1,
photon 2 shows a single-slit diffraction pattern.
Photon 2 does not pass through any slit, and one would not expect it to
show any diffraction.
Another curious feature of this ghost-diffraction is that the fringe width
of the diffraction pattern follows a diffraction formula
$w = {\lambda D\over \epsilon}$, where $\lambda$ is the wavelength of the
photons, $\epsilon$ the width of the slit, and $D$ is a very curious distance
from slit, right through the SPDC source $S$ to the detector D2. Note that
photon 2 does not even travel that much of distance.
The experiment attracted a lot of debate and
research attention \cite{ghostimaging,rubin,zhai,jie,zeil2,twocolor,pravatq,sheebatq,tqrev}.

The double-slit version of this experiment has been understood as a 
combined effect of a virtual double-slit creation for photon 2, and 
{\em quantum erasure} of which-path information\cite{pravatq,sheebatq,tqrev}.
In that analysis the amplitudes of photon 1, to pass through the
two slits, are correlated with certain amplitudes of photon 2. However,
the correlation between the two photons within the narrow region of one slit
had been ignored. In order to explain single-slit ghost diffraction, one would
need to take into account the entanglement between the photons within the
region of a single slit.
A rigorous theoretical analysis of the single-slit ghost experiment would
involve introducing a two-photon entangled state, and evolving it in
time in the presence of single-slit potential. This is a challenging task. 

It has been shown earlier that diffraction of light can be described in terms of
evolution of de Broglie wave-packets\cite{vandegrift,dillon}. Single-slit
diffraction of light in particular, has been described in terms of 
Gaussian wave-packet evolving via Schr\"odinger evolution\cite{zecca}.
In the following we generalise these methods to the case of an
entangled state. We show that the single-slit ghost diffraction can be
explained rather simply using this approach.

\section{Wave-packet analysis}

\subsection{The entangled state}

We start with a generalised EPR state\cite{tqpopper}
\begin{equation}
\Psi(z_1,z_2) = C'\!\int_{-\infty}^\infty dk
e^{-k^2/4\sigma^2}e^{-ikz_2} e^{i kz_1}
e^{-{(z_1+z_2)^2\over 4\Omega^2}}, \label{state}
\end{equation}
where $C'$ is a normalization constant, and $\sigma,\Omega$ are certain
parameters to be explained later. The
state (\ref{state}), unlike the original EPR state\cite{epr}, is well behaved and fully
normalised. In the limit $\sigma,\Omega\to \infty$ the state (\ref{state})
reduces to the original EPR state.

The photons of the pair are assumed to be travelling in opposite directions
along the x-axis, but the entanglement is in the z-direction. In our analysis
we will ignore the dynamics along the x-axis as it does not affect the
entanglement.
We just assume that during evolution for a time $t$, the photon travels a
distance equal to $ct$ in the x-direction. Integration over $k$ can be
performed in (\ref{state}) to obtain:
\begin{equation}
\Psi(z_1,z_2) = \sqrt{ {2\sigma\over \pi\Omega}}
 e^{-(z_1-z_2)^2\sigma^2} e^{-(z_1+z_2)^2/4\Omega^2} .
\label{psi0}
\end{equation}
The uncertainty in position and the wave-vector of the two photons,
along the z-axis, is given by
\begin{eqnarray}
\Delta z_1 = \Delta z_2 = \sqrt{\Omega^2+1/4\sigma^2},~~~~~~~
\Delta k_{1z} = \Delta k_{2z} = 
{1\over 2}\sqrt{\sigma^2 + {1\over 4\Omega^2}}~.
\label{unc}
\end{eqnarray}
From the above it is clear that $\Omega$ and $\sigma$ quantify the position
and momentum (wave-vector) spread of the photons in the z-direction.
Notice however, that for the special case $2\Omega=1/\sigma$, the state
(\ref{psi0}) is factored into a product of two Gaussians for $z_1$
and $z_2$ alone. Thus the state is not entangled for this particular choice of
parameters.

\subsection{Time evolution}

We first lay out our strategy for the time evolution of a photon
wave-packet. Consider the state of a single photon at time $t=0$ as $\psi(z,0)$.
The state at a later time $t$ is given by
\begin{equation}
\psi(z,t) = {1\over 2\pi}\int_{\-\infty}^{\infty}
\exp(ik_zz - i\omega(k_z)t) \tilde{\psi}(k_z,0) dk_z ,
\label{ppsit1}
\end{equation}
where $\tilde{\psi}(k_z,0)$ is the Fourier transform of $\psi(z,0)$ with
respect to $z$. Now photon is approximately travelling in the x-direction,
but can slightly deviate in the z-direction (which allows it to pass through
slits located at different z-positions), so that its true wave-vector will
have a small component in the z-direction too. So
\begin{equation}
\omega(k_z) = c\sqrt{k_x^2 + k_z^2}
\end{equation}
Since the photon is travelling along x-axis by and large, we can write
$k_x \approx k_0$, where $k_0$ is the wavenumber of the photon associated
with its wavelength, $k_0 = 2\pi/\lambda$. The dispersion along z-axis can
then be approximated by
\begin{equation}
\omega(k_z) \approx ck_0 + ck_z^2/2k_0 .
\end{equation}
This is essentially the Fresnel approximation.
Using this approximation, eqn. (\ref{ppsit1}) assumes the form
\begin{equation}
\psi(z,t) = {e^{-ick_0t}\over 2\pi}\int_{-\infty}^{\infty}
\exp(ik_zz - ictk_z^2/2k_0) \tilde{\psi}(k_z,0) dk_z
\label{ppsit2}
\end{equation}

Coming back to our problem of entangled photons,
we assume that after travelling for a time $t_0$, photon 1 reaches the 
slit ($ct_0 = L_2$), and photon 2 travels a distance $L_2$ towards
detector D2.
Using the strategy outlined in the preceding discussion, we can write
the state of the entangled photons after a time $t_0$ as follows:
\begin{eqnarray}
\psi(z_1,z_2,t_0) &=& {e^{-2ick_0t_0}\over 4\pi^2}\int_{-\infty}^{\infty} dk_1
\exp(ik_1z_1 - ict_0k_1^2/2k_0)
\int_{-\infty}^{\infty} dk_2
\exp(ik_2z_2 - ict_0k_2^2/2k_0) \tilde{\psi}(k_1,k_2,0) ,
\label{psit0}
\end{eqnarray}
where $\tilde{\psi}(k_1,k_2,0)$ is the Fourier transform of (\ref{psi0}) with
respect to $z_1,z_2$. After some algebra, the above can be worked out to be
\begin{eqnarray}
\psi(z_1,z_2,t_0) &=& 
C \exp\left[{-(z_1+z_2)^2\over 4\Omega^2+{2i\lambda ct_0/\pi}}\right]
\exp\left[{-(z_1-z_2)^2\over {1\over\sigma^2}+{2i\lambda ct_0/\pi}}\right],
\end{eqnarray}
where $C = \sqrt{2\over\pi}\left[\left\{\Omega^2+\left({\lambda ct_0\over 2\pi\Omega}\right)^2\right\}
\left\{{1\over\sigma^2}+\left({2\sigma\lambda ct_0\over\pi}\right)^2\right\}\right]^{-1/4}$.
The above equation represents the state of the entangled photons just before
photon 1 enters the slit.

\subsection{Effect of slit}

In order to incorporate the effect of the slit of width $\epsilon$ on the
entangled state, 
we just assume that the slit abruptly truncates the state of particle 1 such
that only the part $-\epsilon/2 \le z_1 \le \epsilon/2$ survives.
Consequently we assume this truncated state to be our starting point
after emerging from the slit. After emerging from the slit, photon 1
travels for a time $t$, a distance $L_1$, to reach the detector D1,
and photon 2 travels for
the same time to reach detector D2. Using (\ref{ppsit2}), the time
propagation kernel for the two photons can be written as
\begin{eqnarray}
K_1(z_1,z_1',t) &=& \sqrt{1\over i\lambda ct}
\exp\left[{-\pi(z_1-z_1')^2\over i\lambda ct}\right]\nonumber\\
K_2(z_2,z_2',t) &=& \sqrt{1\over i\lambda ct}
\exp\left[{-\pi(z_2-z_2')^2\over i\lambda ct}\right]
\end{eqnarray}

The two-particle state after a time $t$ (after $t_0$) is given by
\begin{eqnarray}
\psi(z_1,z_2,t) &=& \int_{-\epsilon/2}^{\epsilon/2}dz_1'K_1(z_1,z_1',t)
 \int_{-\infty}^{\infty}dz_2'K_2(z_2,z_2',t)\psi(z_1',z_2',t_0)
\end{eqnarray}
Notice that because of the truncation, the wavefunction above is no
longer normalised. Subsequently we will continue to treat the unnormalised
wavefunction. Integration over $z_2'$ can be performed straightaway to give
\begin{eqnarray}
\psi(z_1,z_2,t) &=& C_r\int_{-\epsilon/2}^{\epsilon/2}
\exp\left[{-\pi(z_1-z_1')^2\over i\lambda ct}\right] 
\exp\left[{-(z_2-z_1'({\Gamma-\gamma\over\Gamma+\gamma}))^2\over
{\Gamma\gamma\over\Gamma+\gamma}+{i\lambda ct\over\pi}}\right]
\exp\left[{-4z_1'^2\over \Gamma+\gamma}\right] dz_1' 
\label{psit}
\end{eqnarray}
where
\begin{equation}
\Gamma = 4\Omega^2+{i2\lambda ct_0\over\pi},~~~~~~
\gamma = {1\over\sigma^2}+{i2\lambda ct_0\over\pi},
\end{equation}
and $C_r = {C\over i\lambda ct}
\left({\Gamma+\gamma\over\pi\Gamma\gamma}+{1\over i\lambda ct}\right)^{-1/2}$.

Assuming the slit is very narrow, we notice that in the integral above,
$|z_1'| \le \epsilon/2 \ll 1$.  Hence we can make the following approximation
\begin{equation}
(z_n - z_1')^2 \approx z_n^2 - 2 z_n z_1',
\end{equation}
where $z_n$ is $z_1$ or $z_2$.
This amounts to neglecting $\epsilon^2$ in comparison to $\epsilon$, and
results in
\begin{eqnarray}
\psi(z_1,z_2,t) &\approx& C_r\int_{-\epsilon/2}^{\epsilon/2}
\exp\left[{i\pi z_1^2-2i\pi z_1z_1'\over \lambda ct}\right] 
\exp\left[{-z_2^2+2z_2z_1'({\Gamma-\gamma\over\Gamma+\gamma})\over
{\Gamma\gamma\over\Gamma+\gamma}+{i\lambda ct\over\pi}}\right]
 dz_1'.
\end{eqnarray}
In (\ref{psit}), the argument of the exponent in the last term is of the
order of $\epsilon^2$, and thus that term has been neglected in writing
the above approximate form.

Integration over $z_1'$ can now be carried out with ease, and one obtains
\begin{eqnarray}
\psi(z_1,z_2,t) &=& C_r \exp\left[{i\pi z_1^2\over\lambda ct}-{z_2^2\over\alpha}\right]
{\sin(\epsilon\pi z_1/2\lambda ct - i\epsilon z_2\beta)\over
\pi z_1/2\lambda ct - iz_2\beta},
\label{finalstate}
\end{eqnarray}
where
$\alpha = {\Gamma\gamma\over\Gamma+\gamma} + {i\lambda ct\over\pi}$ and 
$\beta = {1\over\alpha}\left({\Gamma-\gamma\over\Gamma+\gamma}\right)$.

The probability density of joint detection of photon 2 at $z_2$, and
photon 1 at $z_1$, is given by
\begin{eqnarray}
|\psi(z_1,z_2,t)|^2 &=& |C_r|^2 \left|e^{-z_2^2/\alpha}\right|^2
\left|{\sin(\epsilon\pi z_1/2\lambda ct - i\epsilon z_2\beta)\over
\pi z_1/2\lambda ct - iz_2\beta}\right|^2.
\label{joint}
\end{eqnarray}
Before we come to the results, we recall that the final expression is
understood better in terms of the distance travelled by the photons, rather
than the time for which they evolve. In the following analysis, we 
use $ct_0 = L_2,~ct = L_1$ and $2L_2+L_1 = D$.

\section{Results}

\subsection{Ghost diffraction}

Ghost diffraction arises when photon 2 is detected at D2 in coincidence with
photon 1 being detected at D1 {\em fixed at $z_1=0$}.
The probability density of this happening can be evaluated by putting
$z_1=0$ in (\ref{joint}).
Entanglement between the two photons is good when $\Omega, \sigma \gg 1$.
In this limit we safely assume $\Omega \gg 1/\sigma$. In the approximation 
$2\lambda\L_2/\pi \gg 1/\sigma^2$, this (unnormalised) probability density
assumes the form
\begin{eqnarray}
|\psi(0,z_2,t)|^2 &=& A~{1\over z_2^2}\left[
\sin^2\left({\pi\epsilon z_2\over\lambda D}\right)
\cosh^2\left(\epsilon z_2({\pi\over\sigma\lambda D})^2\right)\right.
\left.+ \cos^2\left({\pi\epsilon z_2\over\lambda D}\right)
\sinh^2\left(\epsilon z_2({\pi\over\sigma\lambda D})^2\right)
\right],
\label{diffr}
\end{eqnarray}
where
\begin{eqnarray}
A = {1\over \sqrt{{\Omega^2+({\lambda L_2\over 2\pi\Omega})^2}}}
\left({D\over 2\pi^2\sigma L_1}\right)
e^{-2({\pi z_2\over\sigma\lambda D})^2}
\end{eqnarray}

\begin{figure}[h]
\centerline{\resizebox{11.0cm}{!}{\includegraphics{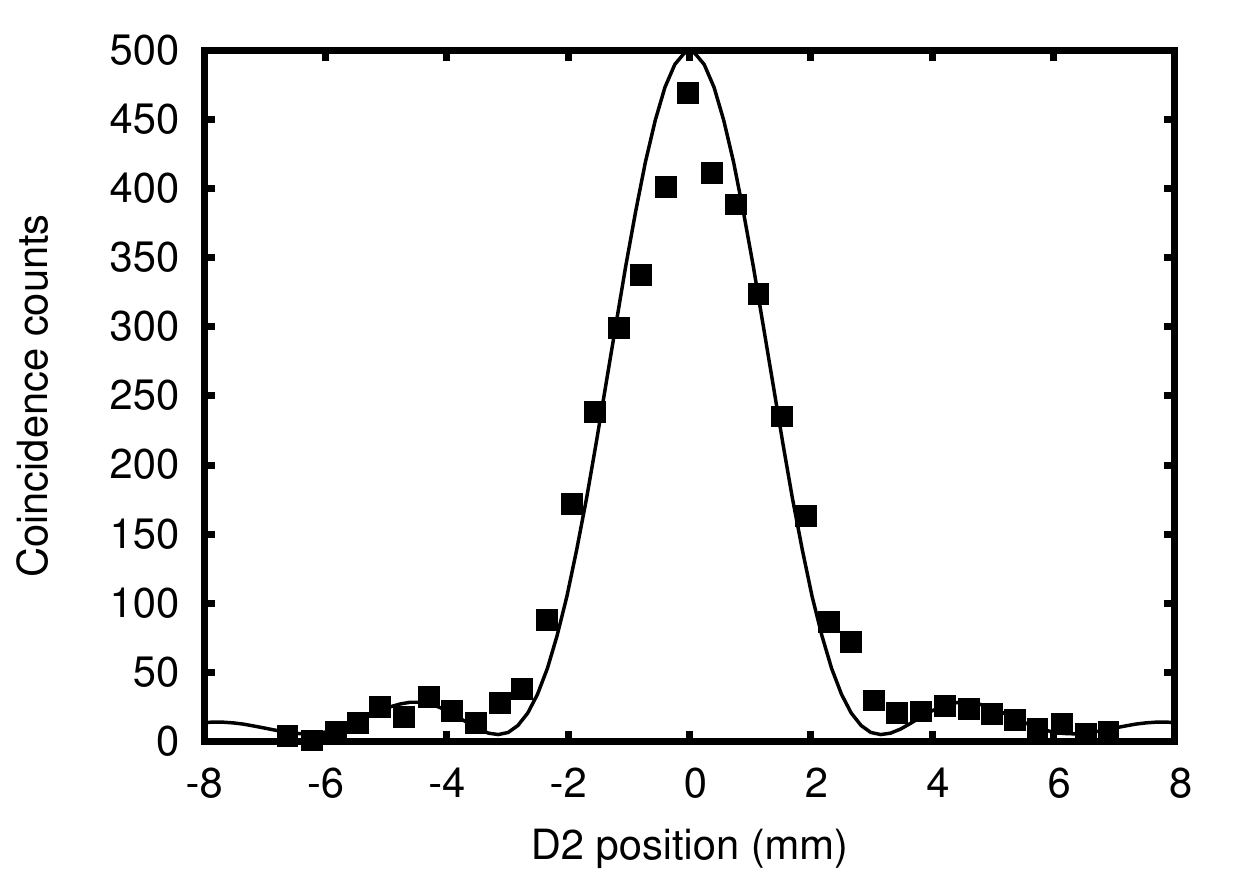}}}
\caption{ Normalised probability density of photon 2 (scaled by a factor of
500), conditioned on photon 1 reaching $z_1=0$, as a function of the position of
detector D2, with D1 fixed at $z_1=0$, for $\lambda=702.2$ nm,
$D = 1.8$ m, $\epsilon=0.4$ mm, and $\sigma=5.0$ mm$^{-1}$.} Squares represent
the experimental data from Strekalov et. al. \cite{ghostexpt}.
\label{plot}
\end{figure}

The probability density, given by (\ref{diffr}), represents a diffraction
pattern for photon 2. In order to compare the theoretical results obtained here
with the experimental results of Strekalov et.al., (\ref{diffr}) has been
plotted after normalising and scaling with 500 counts, and using the other
parameter values of the experiment (see Fig. \ref{plot}). The parameter $\sigma$ has been assigned
an arbitrary value 5 mm$^{-1}$.
The name ghost diffraction is natural for this phenomenon,
as photon 2 does not pass through any real slit, but still shows diffraction.
The distance between the central maximum and the first maximum is given by
\begin{equation}
 w = {\lambda D\over \epsilon}.
\end{equation}
The surprise here is that, taken at face value, the above formula would
represent diffraction of photons of wavelength
$\lambda$ emerging from a slit of width $\epsilon$ and travelling a distance
$D = 2L_2+L_1$ to reach detector D2. However, photons 2 really travel only
a distance $L_1+L_2$.
This is exactly what was seen in the experiment \cite{ghostexpt}.
This quaint feature is the result of entanglement between the two photons.

As a special case, if $2\Omega=1/\sigma$ we find that $\Gamma = \gamma$
and consequently $\beta=0$. In the limit $\beta\to 0$, the pobability density of detecting
photon 2 at a position $z_2$, {\em in coincidence with D1 fixed at $z_1=0$},
is given by
\begin{eqnarray}
|\psi(0,z_2,t)|^2 &=& |C_r|^2 \left|e^{-z_2^2/\alpha}\right|^2 \epsilon^2.
\end{eqnarray}
This is just a Gaussian in $z_2$ and implies that there is no ghost
diffraction, which should not be surprising because for $\beta=0$ the two
photons are disentangled.

Let us try to understand why a coincident count of D2 with a {\em fixed}
D1 is needed for ghost interference to appear. Not doing a coincident count
would amount to integrating over $z_1$, the position of D2, in (\ref{joint}).
It can be easily seen that, for $\beta\ne 0$, integration over $z_1$ would
kill the oscillations in (\ref{joint}), which means no ghost diffraction.

Another way to understand why coincident counting is needed for ghost
diffraction is that if D1 were fixed at $z_1=z_0$ instead of
$z_1=0$, the term
$\left|{\sin(i\epsilon z_2\beta - \pi\epsilon z_0/2\lambda ct)\over
iz_2\beta - \pi z_0/2\lambda ct}\right|^2$ would lead to a slightly shifted
ghost diffraction pattern. This feature has also been experimentally
observed \cite{ghostexpt}. Corresponding to different values of $z_0$,
the diffraction pattern would be shifted by different amounts, and a
sum of all such possibilities would lead to the washing out of the pattern.

\subsection{Missing first order diffraction}

Let us now turn our attention to the issue of missing first order single-slit
diffraction for photon 1 behind the slit. In the first part of the experiment,
photons 1 pass through 
the single slit and are detected by a scanning detector D1. Normally one
would expect single slit diffraction. The authors of the experiment
explain it by saying that the blurring out of the first order interference
fringes is due to the considerably large angular propagation uncertainty
of a single SPDC photon. However, we will show that the real reason lies
in the entanglement of photon 1 with photon 2.

The probability of joint detecttion of the two photons at the two detectors
is given by (\ref{joint}). If one wants to look for first order diffraction
of photon 1, one has to ignore photon 2 or, in other words,
integrate over all values of $z_2$. Clearly, if one were to evaluate
$\int_{-\infty}^{\infty} |\psi(z_1,z_2,t)|^2~dz_2$ using (\ref{joint}),
the oscillations will be gone and it will not yield
a diffraction pattern. Interestingly however, for the case
$2\Omega=1/\sigma$ ($\beta=0$), (\ref{joint}) reduces to
\begin{eqnarray}
|\psi(z_1,z_2,t)|^2 &=& |C_r|^2 \left|e^{-z_2^2/\alpha}\right|^2
\left|{\sin(\epsilon\pi z_1/2\lambda ct )\over
\pi z_1/2\lambda ct}\right|^2.
\end{eqnarray}
The above represents a diffraction pattern for photon 1 without any
coincident counting, since integration over $z_2$ does not effect the
$z_1$ dependent terms. It also says that photon 2 has a Gaussian distribution.
As expected, if the two photons are not entangled,
first order diffraction {\em will be visible} for photon 1, and photon 2
will not show any ghost diffraction. For $\beta\ne 0$,
entanglement will kill any first order diffraction.

Another way to understand the missing first order interference is that
by virtue of entanglement, each path of the photon 1, within the width of
the slit, is correlated with a particular path of of photon 2. By measuring
the position of photon 2, one can, in principle, find out which particular
path photon 1 had taken. If which-path information exists, no interference
between different paths is possible, by virtue of Bohr's complementarity
principle \cite{bohr}.

\section{Conclusion}

We have theoretically analysed the single-slit ghost diffraction experiment
with entangled photons by using Fourier wave-packet evolution. At the heart
of analysis is a generalised EPR state which describes the entangled photons.
The effect of the slit is assumed to be a truncation of the wavefunction of 
the photon
to within the width of the slit. Subsequent evolution is free. This simple
analysis quantitatively reproduces all features of single-slit ghost
diffraction. It explains why the diffraction pattern shifts if the detector
D1 is fixed at a location other than $z_1=0$. It explains the strange distance
that appears in the fringe-width formula of the ghost diffraction, eventhough
the photon never actually travels that distance. This analysis also explains
why first order diffraction behind the single slit is not observed.

\begin{acknowledgement}
Sheeba Shafaq thanks the Centre for Theoretical Physics, JMI, for providing her
the facilities of the centre during the course of this work.
\end{acknowledgement}

\vspace*{-6pt}   %ONLY NECESSARY

\end{document}